\pdfoutput=1
\documentclass[11pt]{article}

\usepackage{coling}

\usepackage{times}
\usepackage{latexsym}

\usepackage[T1]{fontenc}
\usepackage[utf8]{inputenc}

\usepackage{microtype}

\usepackage{inconsolata}

\usepackage{graphicx}
\usepackage{makecell} % 用于创建带有垂直居中的单元格
\usepackage{subcaption}
\title{Learning to Unify Audio, Visual and Text for Audio-Enhanced Multilingual Visual Answer Localization}

\author{Zhibin Wen \\
  Systems Engineering Institute \\ Xi'an Jiaotong University \\ binbin21@stu.xjtu.edu.cn  \\\And
  Bin Li\\
  Shenzhen Institute of Advanced Technology \\ Chinese Academy of Sciences \\ b.li2@siat.ac.cn}

\begin{document}
\maketitle
\begin{abstract}
The goal of Multilingual Visual Answer Localization (MVAL) is to locate a video segment that answers a given multilingual question. Existing methods either focus solely on visual modality or integrate visual and subtitle modalities. However, these methods neglect the audio modality in videos, consequently leading to incomplete input information and poor performance in the MVAL task. In this paper, we propose a unified Audio-Visual-Textual Span Localization (AVTSL) method that incorporates audio modality to augment both visual and textual representations for the MVAL task. Specifically, we integrate features from three modalities and develop three predictors, each tailored to the unique contributions of the fused modalities: an audio-visual predictor, a visual predictor, and a textual predictor. Each predictor generates predictions based on its respective modality. To maintain consistency across the predicted results, we introduce an Audio-Visual-Textual Consistency module. This module utilizes a Dynamic Triangular Loss (DTL) function, allowing each modality's predictor to dynamically learn from the others. This collaborative learning ensures that the model generates consistent and comprehensive answers. Extensive experiments show that our proposed method outperforms several state-of-the-art (SOTA) methods, which demonstrates the effectiveness of the audio modality.
\end{abstract}

\section{Introduction}
With the rapid expansion of the internet, an increasing number of users are turning to online platforms to seek medical advice by posing natural language questions \citep{O’Donnell2023,lim2022}. Current online platforms typically fall into two categories: those that provide textual answers, which may be difficult for users to interpret, and those that offer visual answers, which are generally more intuitive and easier to follow \citep{tang2021}. However, the retrieved videos often contain substantial amounts of information irrelevant to the user's query \citep{moon2023}, which significantly hinders the efficiency of information retrieval \citep{zhang2023}. In response to this challenge, the task of Visual Answer Localization (VAL) has been introduced \citep{weng2023}.

\begin{figure}[t]
  \includegraphics[width=\columnwidth]{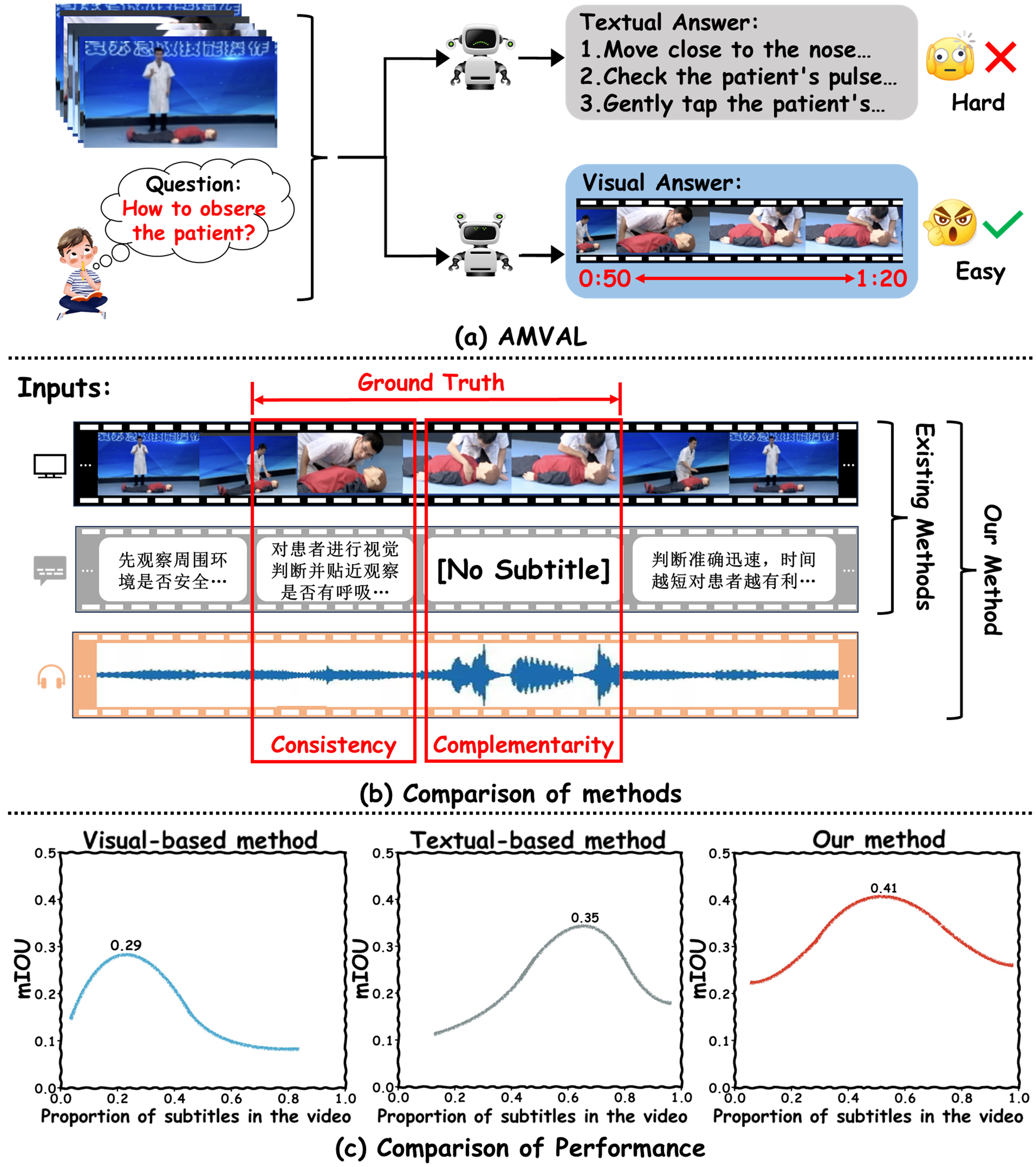}
  \caption{(a) Overview of the audio-enhanced multilingual video answer localization task. (b) Difference between existing methods and our method. (c) Performance comparison diagram of visual-based, textual-based, and our method.}
  \label{fig:introduction}
\end{figure}

Existing VAL approaches can be broadly categorized into visual-based \citep{tang2021frame,chen2020rethinking} and textual-based methods \citep{li2023,weng2023,li2024}. Visual-based methods are effective in scenarios where subtitle text is sparse, but their performance tends to degrade significantly in other contexts. In contrast, textual-based methods excel when abundant subtitle text is available, as the semantic similarity between the question and subtitle is typically greater than between the question and the video \citep{li2024}. However, these methods often overlook audio, which plays a crucial role in complementing both visual and textual modalities. There is inherent consistency and complementarity among these modalities \citep{chen2023curriculum}, and harnessing this synergy can enhance both visual and textual modalities by integrating information from the audio. Incorporating audio thus addresses the performance limitations in VAL, particularly in video segments lacking subtitles \citep{liu2022umt,chen2020learning,sun2024tr}.

To this end, we study the Audio-enhanced Multilingual Visual Answer Localization (AMVAL) which aims to locate video segments that answer a user's natural language question, in either Chinese or English. By providing video segments with verbal explanations for medical guidance, this approach not only facilitates the learning of specific actions but also helps bridge language barriers, making the content accessible to people who speak different languages \citep{macedonia2011body,diamond2020}. However, a significant challenge lies in effectively integrating the three modalities and fully utilizing their individual strengths to tackle the AMVAL task.

To address this challenge, we propose a unified Audio-Visual-Textual Span Localization (AVTSL) method for AMVAL, aimed at reducing cross-modal discrepancies and improving the accuracy of span localization by integrating audio modality. We designed a network architecture with three modality channels (audio, visual, and textual) to fully leverage the semantic information from each modality in the video, addressing the limitations of single-modality methods. Each channel is equipped with a corresponding predictor: an audio-visual predictor, a visual predictor, and a textual predictor. During joint training, distinct objectives are assigned to each predictor, enabling them to leverage the unique strengths of their respective modalities. To improve modality integration, we introduce an Audio-Visual-Textual consistency module, which employs a Dynamic Triangular Loss (DTL) function based on Intersection over Union (IoU). This loss function aligns the modalities by minimizing the discrepancies between each predictor's output and the target answer, as well as between the outputs of the other two predictors. Our approach promotes mutual learning among the predictors to achieve consistent and cohesive multimodal representations.

Our contributions are as follows: (1) We study the AMVAL and propose the AVTSL method, which is the first to introduce the audio modality for the AMVAL; (2) We designed an Audio-Visual-Text Consistency module, which leverages the consistency and complementarity between different modalities using the DTL loss function; (3) We conducted extensive experiments to demonstrate the effectiveness of the AVTSL method, where our method outperformed other state-of-the-art (SOTA) methods by incorporating the audio modality.

\section{Related Work}
\textbf{Video Understanding.} Video understanding is a challenging task that requires comprehension of complex semantic information \citep{diba2019holistic}. Audio is a typical modality in video understanding tasks that has been studied for many years. Some datasets \citep{chen2023valor,yang2022avqa} focus on the incorporation of the audio modality. A recent work \citep{li2023progressive} uses an audio-guided visual attention module to link audio with relevant spatial regions. \citep{wang2023tiva} propose a multimodal knowledge graph that integrates text, image, video, and audio to support diverse downstream tasks. \citep{li2024object} focus on aligning audio and visual modalities on key question words and employ a modality-conditioned module to emphasize relevant audio segments or visual objects. \citep{chen2023curriculum} introduce a novel curriculum-based denoising strategy that adaptively assesses sample difficulty to gauge noise intensity in a self-aware manner. \citep{ibrahimi2023audio} uses two independent cross-modal attention blocks to allow the text to separately focus on audio and video representations. The audio modality has been widely utilized in various video understanding tasks, such as video question answering and temporal sentence grounding in video, with its effectiveness well demonstrated. Inspired by these works, our work is the first to incorporate the audio modality into the AMVAL task. 

\begin{figure*}[t]
  \includegraphics[width=\linewidth]{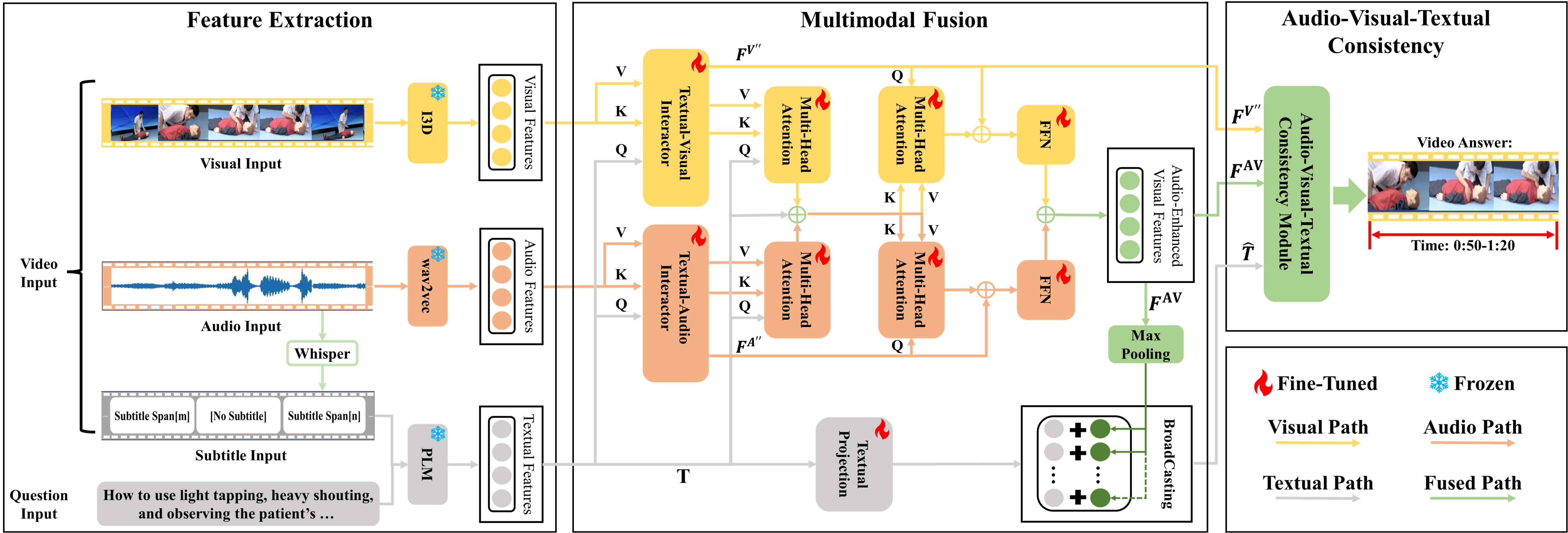} 
  \caption {Overview of the proposed unified Audio-Visual-Textual Span Localization (AVTSL) method}
  \label{AVTSL}
\end{figure*}

\textbf{Multimodal Fusion.} Recent works \citep{chen2019motion,liu2022umt} suggest that leveraging motion and audio can improve performance, but they fall short by merely concatenating features from multiple modalities without considering their relative importance or interactions, leaving their full potential untapped. A few studies \citep{hori2017attention,jin2019low} assign importance weights to individual modalities using cross-modal attention in the encoder, but they still lack explicit handling of modality interactions. \citep{lan2023} proposes a deep module for learning shared information to enhance representation in multimodal sentiment analysis. \citep{wang2023tedt} introduces a transformer-based multimodal encoding-decoding network to address the challenges of multimodal sentiment analysis, focusing on the influence of individual modal data and the low quality of nonverbal features. \citep{chen2020learning} proposes the Pairwise Modality Interaction (PMI) method that uniquely captures pairwise modality interactions at both sequence and channel levels. This method preserves the temporal dimension and fuses interaction results with importance weights for added explainability. Inspired by these modality fusion methods, we designed three multimodal information fusion channels: audio-textual, visual-textual, and audio-visual-textual.

\section{Task Definition}
Given an untrimmed video \(V\) and a text question \(Q = \{q_m\}^{L_q}_{m=1}\), where \(Q\) can be in either Chinese or English and \(L_q\) is the length of the question. \(V = \{F^v,F^a\}\) is representing the visual modality \(F^v = \{f^v_t\}_{t=1}^{T_v}\) and audio modality \(F^a = \{f^a_t\}_{t=1}^{T_a}\) in the input video \(V\), where \(T_v\) and \(T_a\) are the number of frames and duration in seconds of video \(V\). Corresponding subtitle text \(S = \{F^s\}\) can be extracted from the audio modality, where \(F^s = \{f^s_j\}_{j=1}^{L_s}\) denotes subtitle of each video, \(L_s\) is the subtitle span length. The goal of the AMVAL task is to predict the start and end timestamps of the video segment that answers the given question \(Q\), whose ground truth annotations are \(y = (y^s,y^e)\).

\section{Proposed Method}
Figure~\ref{AVTSL} gives an overview of our proposed AVTSL, which is composed of three key components: Feature Extraction, Multimodal Fusion, and Audio-Visual-Textual Consistency.

\begin{figure*}[t]
  \includegraphics[width=\linewidth]{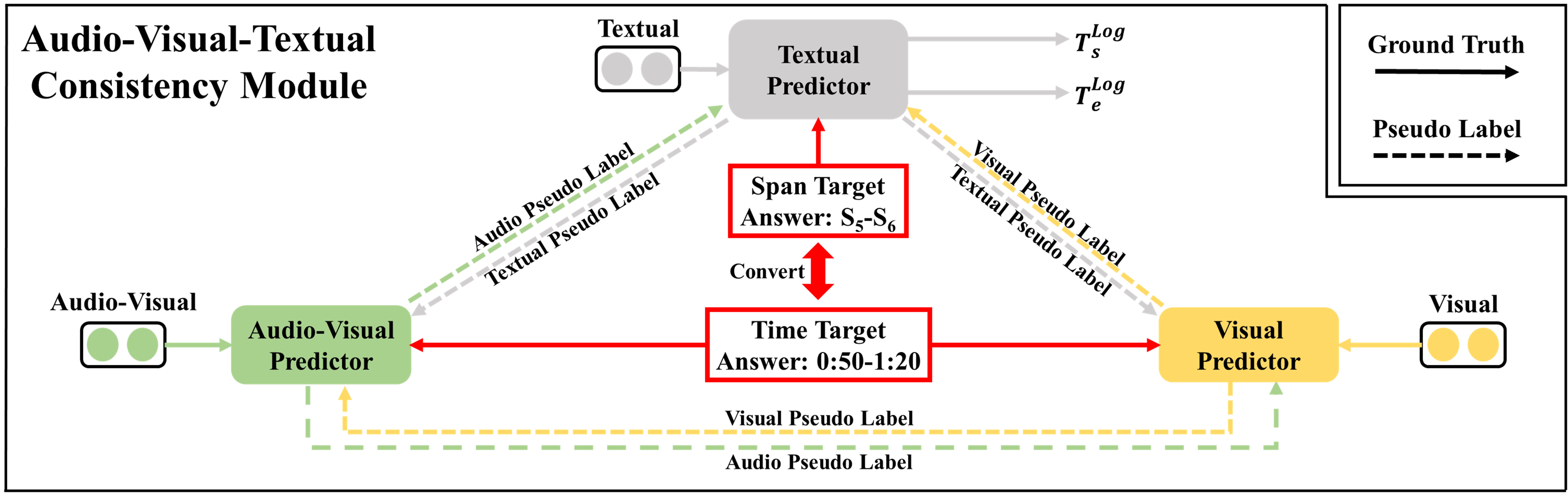} 
  \caption {Overview of the proposed Audio-Visual-Textual Consistency module}
  \label{AVTC}
\end{figure*}

\subsection{Feature Extraction}
For each video \(V\), we extract visual frames (16 frames per second) and then obtain the corresponding RGB visual features \({F^v} =\{v_k\}^{V_n}_{k=1} \in \mathbf{R}^{{V_n}\times d_v}\) using Inflated 3D ConvNet (I3D) pre-trained on the Kinetics dataset \citep{carreira2017quo}, where \(V_n\) is the number of extracted features and \(d_v\) is the dimension of the
visual features. We employ the wav2vec pre-trained  model \citep{baevski2020wav2vec} with audio projection to extract audio features \({F^a} =\{a_i\}^{A_n}_{i=1} \in \mathbf{R}^{{A_n}\times d_a}\)from the video \(V\), where \(A_n\) is the number of extracted features and \(d_a\) is the dimension of the audio features. For the text component, we first extract the raw subtitle text \(S = \{s_q\}_{q=1}^n\) from the audio using the general-purpose speech recognition model (Whisper) \citep{radford2023robust}, where \(n\) is the subtitle span
length. Next, we concatenate the text question \(Q\) with the video's raw subtitle text \(S\) to form \({T}' = [Q,s_i],i \in [1,n]\). Finally, after tokenizing the concatenated text \(T'\), we use the DeBERTa pre-trained model with textual projection to obtain the textual features \({T} = \{t_i\}_{i=1}^{T_n} \in \mathbf{R}^{{T_n}\times d_t}\), where the \(T_n\) is the length of tokenized concatenate text and the \(d_t\) is the dimension of textual features.

\subsection{Multimodal Fusion}
\textbf{Textual-Visual/Audio Interactor.} After extracting visual (\({F^v}\)), audio (\({F^a}\)), and textual \({T}\) features, we employ a context query attention(CQA) model, inspired by the work \citep{zhang-etal-2020-span}, to capture interactions between the textual, visual and audio modalities. As shown in Figure~\ref{AVTSL}, we employed two modality interactors (Textual-Visual/Audio Interactor) with same architectures, capable of facilitating interactions between feature sequences from two modalities, denoted by \(T\) and \(F^m\), where \(m \in \{v, a\}\). 

Modality interaction is divided into two parts: (1) context-to-query and query-to-context processes; (2) context query concatenation module. First, we calculate two attention weights as:
\begin{equation}
  \label{eq:1}
\mathcal{S}^{c2q} = \mathcal{S}_r \cdot T,
\mathcal{S}^{q2c} =  \mathcal{S}_c \cdot \mathcal{S}_r^T \cdot F^m
\end{equation}
where \(\mathcal{S}^{c2q} \in \mathbf{R}^{T_m \times d_t}\) and \(\mathcal{S}^{q2c} \in \mathbf{R}^{T_m \times (d_m)}\) denotes context-to-query and query-to-context processes.  \(\mathcal{S}_r \in  \mathbf{R}^{T_m \times T_n}\) and  \(\mathcal{S}_c \in  \mathbf{R}^{T_m \times T_n}\) denotes the row-wise and column-wise normalization of \(\mathcal{S}\) by SoftMax. And then we obtain the output of CQA module as:
\begin{equation}
  \label{eq:2}
{{F}^m}' = FFN([F^m; \mathcal{S}^{c2q}; F^m \odot \mathcal{S}^{c2q}; F^m \odot \mathcal{S}^{q2c}])
\end{equation}
where FFN is a feed-forward network; \(\odot\) is Hadamard product.

Second, we capture deeper semantic interactions between textual and visual/audio modalities through the context query concatenation module, which is computed as:
\begin{equation}
  \label{eq:3}
{{F}^m}'' = Conv1d(Concat[attn({{F}^m}', T); T])
\end{equation}
where \(attn\) denotes the attention layer. Finally, we obtain the Textual-Visual/Audio fused features that emphasizes the audio and visual segments that are semantically relevant to the text query.

\textbf{Audio-Visual-Textual Interaction.} As shown in Figure~\ref{AVTSL}, after obtaining the feature vectors \({F^{V}}''\) and \({F^{A}}''\), we further apply a multi-head attention mechanism \citep{chen2023curriculum} to derive the fused representation of the three modalities. First, we use the text features to query the relevant visual and audio features, then apply a residual connection with the text features.
\begin{equation}
  \label{eq:4}
  \begin{array}{lcr}
F^{T_1} = MHA(q=T, k={F^{V}}'', v={F^{V}}'')\\
F^{T_2} = MHA(q=T, k={F^{A}}'', v={F^{A}}'')
\end{array}
\end{equation}
\begin{equation}
  \label{eq:5}
T' = F^{T_1} + F^{T_2} + T
\end{equation}
where MHA is  multi-head attention layer.

Next, we propagate \(T'\) to the audio and visual features, using 
visual/audio features as the query and added the outputs of mutil-head attention module:
\begin{equation}
  \label{eq:6}
 \begin{array}{lcr}
{F^{V}}''' = MHA({F^{V}}'', T', T') + {F^{V}}''\\
{F^{A}}''' = MHA({F^{A}}'', T', T') + {F^{V}}''
\end{array}
\end{equation}
\begin{equation}
  \label{eq:7}
F^{AV} = {F^{V}}''' + {F^{A}}'''
\end{equation}
Finally, we obtain the Auido-Enhanced Visual features \(F^{AV}\).
For the textual features, we embed the max-pooled Auido-Enhanced Visual features \(F^{AV}\) into each token of textual features \(F\) based on broadcast mechanism:
\begin{equation}
  \label{eq:8}
\hat{F}^{AV} = MaxPool(F^{AV})
\end{equation}
\begin{equation}
  \label{eq:9}
\hat{T} = \{\hat{F}^{AV} + T_i\}_{i=1}^N
\end{equation}

\subsection{Audio-Visual-Textual Consistency}
\textbf{Three Predictors.} As shown in Figure~\ref{AVTC}, we designed a tri-modal predictor architecture corresponding to three types of modality interaction data features: audio-visual, visual, and textual. The audio-visual and visual span predictors use the same structure, both using two unidirectional LSTM networks(\(LSTM_s\) and \(LSTM_e\)) and two FFN layers(\(FFN_s\) and \(FFN_e\)) to predict the start and end time point logits of the answer segment.
\begin{equation}
  \label{eq:10}
 \begin{array}{lcr}
AV^{Log}_s = FFN_s^{AV}(LSTM_s({F}^{AV}))\\
AV^{Log}_e = FFN_e^{AV}(LSTM_e({F}^{AV}))
\end{array}
\end{equation}
\begin{equation}
  \label{eq:11}
 \begin{array}{lcr}
V^{Log}_s = FFN_s^{V}(LSTM_s({F^{V}}''))\\
V^{Log}_e = FFN_e^{V}(LSTM_e({F^{V}}''))
\end{array}
\end{equation}
The textual predictor differs from the audio-visual and visual span predictors described above, utilizing only two FFN layers(\(FFN_s\) and \(FFN_e\)) to predict the start and end span of the subtitle segments corresponding to the answer.
\begin{equation}
  \label{eq:12}
 \begin{array}{lcr}
T^{Log}_s = FFN_s^{T}(\hat{T})\\
T^{Log}_e = FFN_e^{T}(\hat{T})
\end{array}
\end{equation}

One of the training objectives of the audio-visual/visual predictor is to predict the start and end timestamps of the answer video segment. Since the answer video segment may contain parts without subtitles, the start and end positions of the subtitle segments predicted by the textual predictor may not align with the timestamps of the answer video segment. Therefore, we use a Time-Span Mapping Table \(\mathcal{TSM}\) to convert this:
\begin{equation}
  \label{eq:13}
\hat{T}_t^M = Argmin(M_t - \mathcal{TSM}(T_i))
\end{equation}
\begin{equation}
  \label{eq:14}
\hat{M}_t = Argmin(T_t - \mathcal{TSM}(M_i))
\end{equation}
where  \(M \in (V, AV)\) denotes two types of modality: visual and audio-visual. \(t \in (s,e)\) denotes start and end point, \(\mathcal{TSM}\) stores a one-to-one mapping between video timestamps and subtitle spans.

\textbf{Dynamic Triangular Loss (DTL) Function.} To maintain consistency among the three modalities and leverage the strengths of each, we train the three predictors jointly. We define two types of training objectives: the first is to ensure that each predictor's output closely matches the ground truth; the second is to ensure that each predictor's output remains consistent with the other two predictors, allowing them to benefit from the strengths of the other modalities to compensate for their own weaknesses. Therefore, we set the predictions of each predictor as pseudo-labels for the other two predictors.

First, we calculate the loss for each of the three predictors based on the labeled answer segments.
\begin{equation}
  \label{eq:15}
\mathcal{L}_1^{AV} = CE(AV^{Log}_s, AV_s) + CE(AV^{Log}_e, AV_e)
\end{equation}
\begin{equation}
  \label{eq:16}
\mathcal{L}_1^V = CE(V^{Log}_s, V_s) + CE(V^{Log}_e, V_e)
\end{equation}
\begin{equation}
  \label{eq:17}
\mathcal{L}_1^T = CE(T^{Log}_s, T_s) + CE(T^{Log}_e, T_e)
\end{equation}
where \(CE(\cdot)\) denotes Cross-Entropy function. \([AV_s,AV_e]\), \([V_s,V_e]\) denotes the start and end timestamps of the target answer segment and \(AV_s = V_s = y^s, AV_e = V_e = y^e\). \(T_s\) and \(T_e\) are obtained by converting \(V_s\) and \(V_e\) using the Time-Span Mapping Table \(\mathcal{TSM}\).

Secondly, we calculate the loss between audio-visual predictor and the outputs of the other two predictors:
\begin{equation}
  \label{eq:18}
 \begin{array}{lcr}
\mathcal{L}_2^{AV-T}& = CE(AV^{Log}_s, sg(\widehat{AV}_s)) \\
&+ CE(AV^{Log}_e, sg(\widehat{AV}_e))
\end{array}
\end{equation}
\begin{equation}
  \label{eq:19}
 \begin{array}{lcr}
\mathcal{L}_2^{AV-V} &= CE(AV^{Log}_s, sg(V^{Log}_s)) \\
&+ CE(AV^{Log}_e, sg(V^{Log}_e))
\end{array}
\end{equation}

\begin{table*}
  \centering
  \begin{tabular}{lllllll}
    \hline
    \textbf{Method}
    & \textbf{Publication}
    & \textbf{Src.}
    & \textbf{IoU=0.3} 
    & \textbf{IoU=0.5} 
    & \textbf{IoU=0.7}
    & \textbf{mIoU}\\
    \hline
    Random Pick &- &- &5.71  &4.65  &3.58  &3.97 \\
    DEBUG \citep{lu-etal-2019-debug} &EMNLP & V  
    &25.85  &13.08  &6.74  &16.32                  \\
    GDP \citep{chen2020rethinking} &AAAI  & V  
    &27.34  &15.21  &6.82  &16.87                  \\
    ACRM \citep{tang2021frame} &TMM  & V  
    &23.65  &14.38  &5.26  &15.83                  \\
    MutualSL \citep{weng2023} &ICASSP  & VT  
    &40.55  &29.11  &14.54  &28.98                  \\
    FMALG \citep{cheng2023unified}  &NLPCC & VT  
    &40.99  &28.63  &15.44  &29.77                  \\
    OCR-LLM \citep{HuanZhang2024impro} &NLPCC & VT  
    &50.88  &35.42  &20.54  &36.37                  \\
    VPTSL \citep{li2024} &TPAMI & VT  
    &51.74  &34.03  &17.01  &36.32                  \\
    PMI-LOC \citep{chen2020learning} &ECCV & AV 
    &32.05  &14.28  &6.78  &21.84                    \\
    ADPN \citep{chen2023curriculum} &ACM MM  & AV 
    &29.97  &15.22  &7.85  &22.22                    \\
    ADPN \citep{chen2023curriculum} &ACM MM  & AVT 
    &32.64  &17.36  &9.38  &24.43                    \\
    TR-DETR \citep{sun2024tr} &AAAI & AVT 
    &31.86  &19.94  &10.25  &26.37                    \\
    AVTSL  &- & AVT 
    &\textbf{58.08}  &\textbf{41.02} 
    &\textbf{29.04}  &\textbf{41.75}                 \\
    \hline
  \end{tabular}
  \caption{
    Performance (\%) comparison on MMIVQA dataset. The best score in each column is highlighted in bold. "V" represents the input of only the visual modality. "VT" represents the input of visual-textual modalities. "AV" represents the input of audio-visual modalities. and "AVT" represents the input of audio-visual-textual modalities.}
  \label{overallresult}
\end{table*}

Similarly, we can then compute the loss between the visual predictor and the outputs of the other two predictors:
\begin{equation}
  \label{eq:20}
 \begin{array}{lcr}
\mathcal{L}_2^{V-T}& = CE(V^{Log}_s, sg(\widehat{V}_s)) \\
&+ CE(V^{Log}_e, sg(\widehat{V}_e))
\end{array}
\end{equation}
\begin{equation}
  \label{eq:21}
 \begin{array}{lcr}
\mathcal{L}_2^{V-AV} &= CE(V^{Log}_s, sg(AV^{Log}_s)) \\
&+ CE(V^{Log}_e, sg(AV^{Log}_e))
\end{array}
\end{equation}
The loss between the textual predictor and the outputs of the other two predictors:
\begin{equation}
  \label{eq:22}
 \begin{array}{lcr}
\mathcal{L}_2^{T-AV}& = CE(T^{Log}_s, sg(\widehat{T}^{AV}_s)) \\
&+ CE(T^{Log}_e, sg(\widehat{T}^{AV}_e))
\end{array}
\end{equation}
\begin{equation}
  \label{eq:23}
 \begin{array}{lcr}
\mathcal{L}_2^{T-V}& = CE(T^{Log}_s, sg(\widehat{T}^V_s)) \\
&+ CE(T^{Log}_e, sg(\widehat{T}^V_e))
\end{array}
\end{equation}

At the early stages of training, the output differences between predictors are large and unstable. To enhance training stability, the weight of the second part of the loss (consistency between predictors) should be reduced, with more focus on the difference between each predictor and the ground truth. Therefore, we adjust the weight of the second part of the loss based on the intersection-over-union (IoU) \citep{rodriguez2020proposal} of the outputs from different modality predictors:
\begin{equation}
  \label{eq:24}
\lambda_{M-N} = IOU([\hat{M}_s ,\hat{M} _e],[M_s,M_e])
\end{equation}
where \(M\) denotes the output of the \(M\) modality predictor, and \(\hat{M}_s\) represents the output of the \(N\) modality predictor after being converted by the Time-Span Mapping Table \(\mathcal{TSM}\). After this, we can compute the second part of loss function as :
\begin{equation}
  \label{eq:25}
\mathcal{L}_2 = \sum{\lambda_{M-N}\mathcal{L}_2^{M-N}}
\end{equation}
where \(M, N \in (AV, V, T)\) and \(M \ne N\) denotes two different types of modalities.

Finally, our loss function is:
\begin{equation}
  \label{eq:26}
\mathcal{L} = \mathcal{L}_1^{AV} + \mathcal{L}_1^{V} + \mathcal{L}_1^{T} + \mathcal{L}_2
\end{equation}

\section{Experiments}
\subsection{Datasets and Settings}
We conducted our experiments on a public dataset for VAL: MMIVQA dataset \citep{li2024overview}. The dataset consists of complete medical instructional videos with corresponding questions (in Chinese and English) and video answer timestamps. Each video may contain multiple question-answer pairs, with identical questions linked to a single answer. The dataset is divided into 3768, 334, and 288 pairs for training, validation, and testing, respectively.

Following previous work \citep{chen2023curriculum,li2024,weng2023}, we use "R@n, IoU=\(m\)" (\(n=1\) and \(\mu \in 0.3, 0.5, 0.7\)) and "mIOU" as our evaluation metric. "R@n, IoU=m" denotes the Intersection-over-Union (IoU) of the predicted video answer span compared to the ground truth, measuring the overlap that exceeds a threshold \(m\) in the top-n retrieved moments. The “mIoU” is the average IoU between the predictions and the ground truth across all samples.

We initialize audio features using the pre-trained wav2vec model \citep{baevski2020wav2vec} and extract visual features with the I3D network \citep{carreira2017quo}. Subtitles are generated by the Whisper model \citep{radford2023robust}, while both questions and subtitles are encoded using DeBERTa-v2 \citep{fengshenbang}. The AdamW optimizer \citep{loshchilov2017decoupled} is used with a learning rate of 8e-6, and feature dimensions set to 1024 for all modalities. The model is trained for 15 epochs with a batch size of 1 on an NVIDIA RTX 8000 GPU, with each experiment repeated three times to reduce random errors.

\begin{table*}
  \centering
  \begin{tabular}{lcccc}
    \hline
    \textbf{Method}         
    & \textbf{IoU=0.3} 
    & \textbf{IoU=0.5} 
    & \textbf{IoU=0.7}
    & \textbf{mIoU}  \\
    \hline
    (1) AVTSL(VP) w/o DTL &17.52  &12.05  &8.97  &15.93 \\
    (2) AVTSL(AP) w/o DTL &20.33  &14.74  &10.26  &17.54 \\
    (3) AVTSL(TP) w/o DTL &46.82  &33.62  &19.43  &34.25 \\
    (4) AVTSL(VP) w/o Audio  &25.63  &13.69  &9.32  &21.18 \\
    (5) AVTSL(TP) w/o Audio  &47.57  &34.72  &19.10  &35.27 \\
    (6) AVTSL(VP)  &28.97  &15.58  &10.62  &23.87 \\
    (7) AVTSL(AP)  &30.54  &17.92  &11.07  &25.39 \\
    (8) AVTSL(TP)  &\textbf{58.08}  &\textbf{41.02} 
               &\textbf{29.04}  &\textbf{41.75}  \\
    \hline
  \end{tabular}
  \caption{Ablation Studies of the proposed AVTSL on MMIVQA dataset.}
  \label{ablationresult}
\end{table*}

\subsection{Overall Performance}
In Table~\ref{overallresult}, we compare the performance of our AVTSL model with several strong baselines: 1) \textbf{Random Pick:} A baseline that randomly selects the answer span; 2) \textbf{Visual methods:} DEBUG \citep{lu-etal-2019-debug}, GDP \citep{chen2020rethinking}, and ACRM \citep{tang2021frame}; 3) \textbf{Visual-Textual Fusion methods:} MutualSL \citep{weng2023}, FMALG \citep{cheng2023unified}, OCR-LLM \citep{HuanZhang2024impro}, and VPTSL \citep{li2024}; 4) \textbf{Audio-Visual Fusion methods:} PMI-LOC \citep{chen2020learning} and ADPN (without subtitle) \citep{chen2023curriculum}. 5) \textbf{Audio-Visual-Textual Fusion methods:} ADPN (with subtitle) \citep{chen2023curriculum} and TR-DETR \citep{sun2024tr}.

Our method outperforms strong baselines across all metrics, as shown in Table~\ref{overallresult}. The random selection method performs the worst, demonstrating the challenge of the MMIVQA task. Visual-based methods improve significantly over random selection, indicating that visual features contain rich semantic information. Textual-based methods, however, surpass visual-based ones by nearly 100\% across most metrics, particularly in IoU=0.7, where the semantic similarity between the question and subtitle proves stronger than with video content. This suggests that textual features are more effective for AMVAL. Audio-visual method improves upon visual-only methods, suggesting that incorporating the audio modality enhances visual modality. However, the gains from audio remain smaller than those from textual-based approaches, emphasizing subtitle text as the primary source of information for video answer localization.

This observation is further validated by the experimental results of our AVTSL method. Compared to the visual-based GDP method, AVTSL improves by 31\%, 26\%, 22\%, and 25\% in IoU=0.3, IoU=0.5, IoU=0.7, and mIoU metrics, respectively. Against the textual-based VPTSL method, it shows gains of 7\%, 7\%, 9\%, and 5\% in the same metrics, and compared to the audio-visual PMI-LOC method, improvements are 26\%, 27\%, 23\%, and 20\%. These results demonstrate that the advantage of integrating audio, visual, and textual modalities for the AMVAL task, as their consistency enhances performance. Furthermore, incorporating all three modalities leads to superior performance compared to using any single modality, showing their complementary nature. Notably, under same video modality conditions, our AVTSL shows significantly better performance compared to the ADPN and TR-DETR. This is because ADPN and TR-DETR rely only on audio-visual or visual predictors, whereas we incorporates predictors from all three modalities, with the textual predictor playing a key role.

\begin{figure*}[t]
  \includegraphics[width=\linewidth]{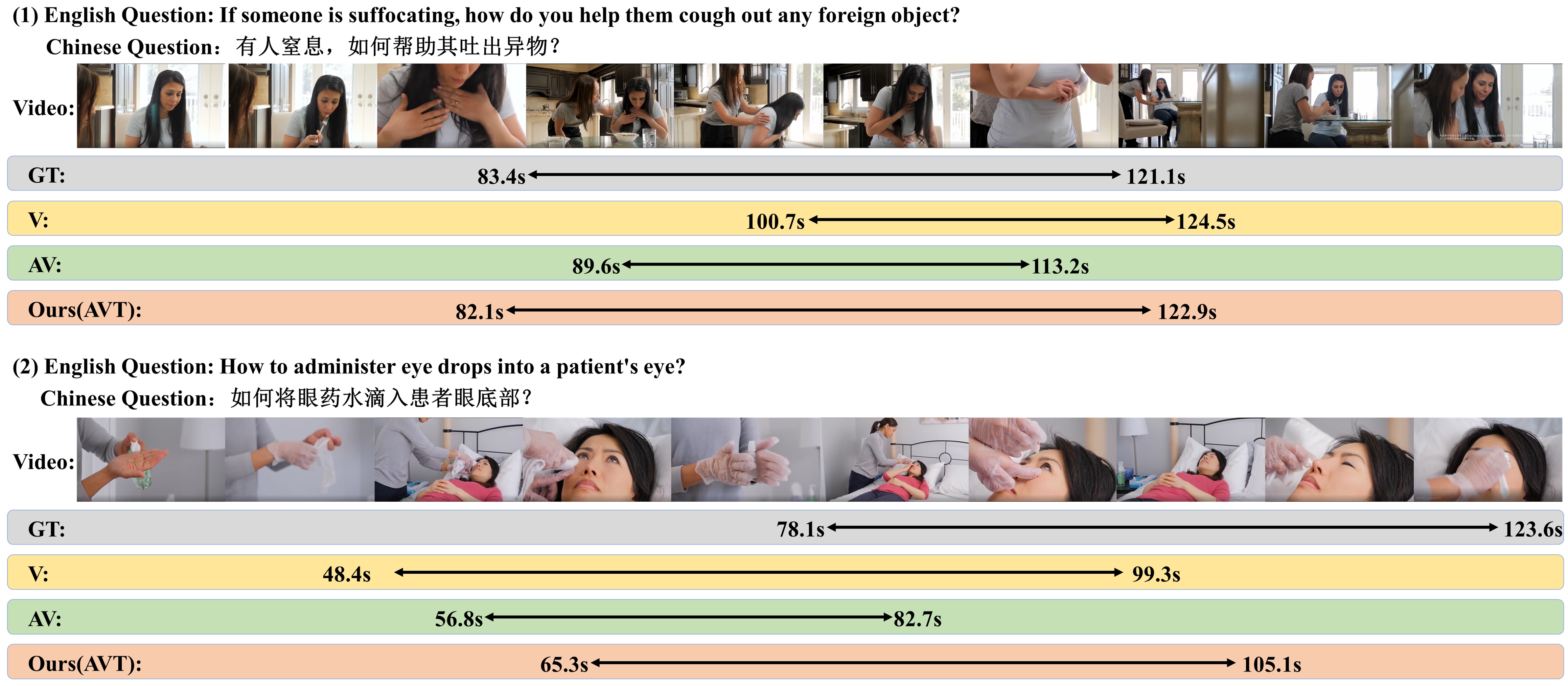}
  \caption {Sample results of different methods performed on the MMIVQA. "GT" represents the ground truth. "V" represents the input of only the visual modality. "AV" represents the input of audio-visual modalities. and "AVT" represents the input of audio-visual-textual modalities.}
  \label{casestudy}
\end{figure*}

\begin{figure}[t]
	\centering
	\begin{subfigure}{0.48\columnwidth}
		\includegraphics[width=\linewidth]{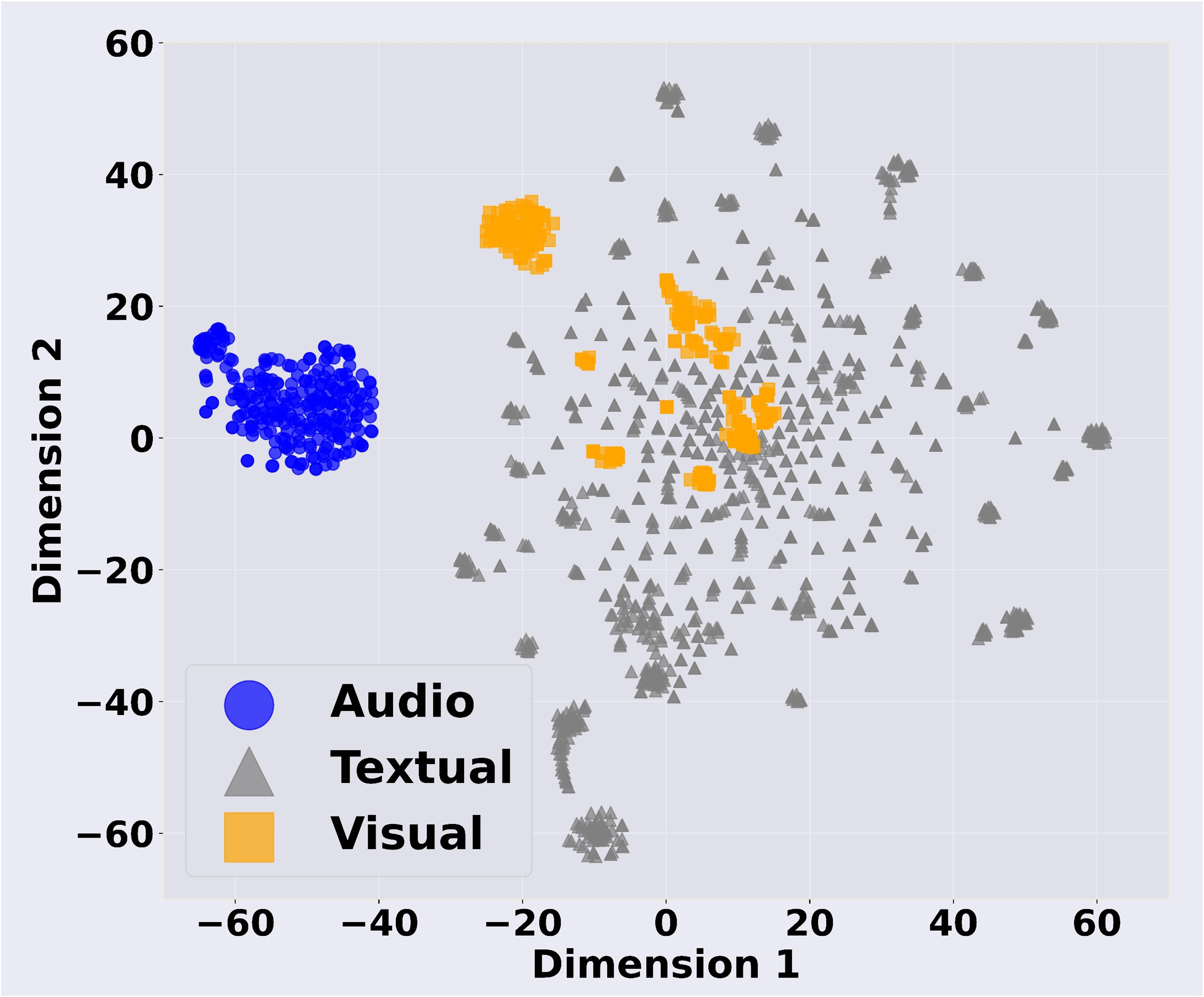}
		\caption{Before interaction}
		\label{fig:subfig1}
	\end{subfigure}
	\hfill
	\begin{subfigure}{0.48\columnwidth}
		\includegraphics[width=\linewidth]{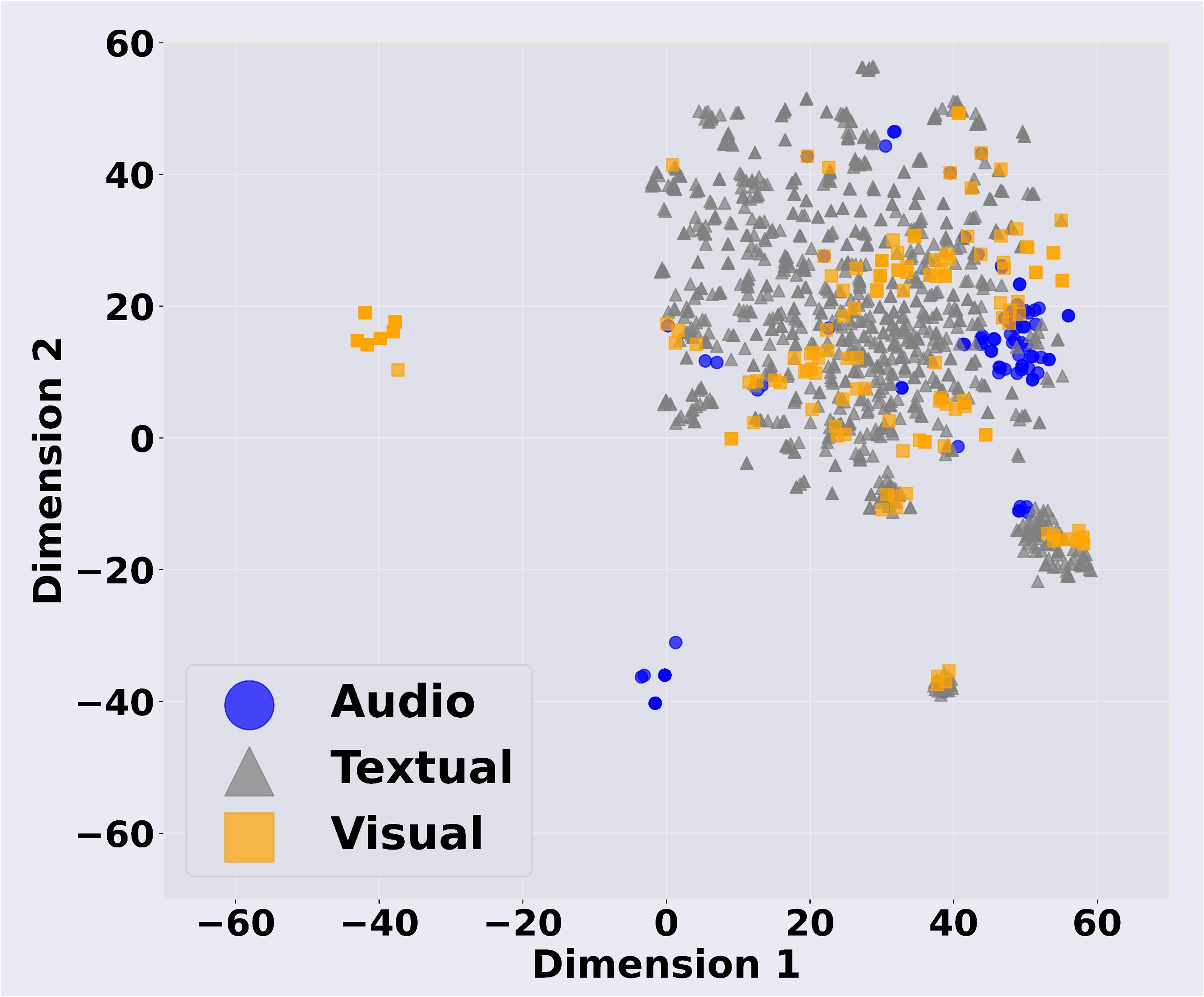}
		\caption{After interaction}
		\label{fig:subfig2}
	\end{subfigure}
	\caption{t-SNE visualization of feature vectors from the three modalities for the same video: (a) before modality interaction, (b) after modality interaction.}
	\label{tsne}
\end{figure}

\subsection{Ablation Study}
We perform ablation studies to assess the contribution of each component in our AVTSL. The implementation details are as follows: 1) "VP," "AP," and "TP" denote the outputs of the visual, audio-visual, and textual predictors, respectively. 2) "w/o DTL" denotes to the model without DTL function. 3) "w/o Audio" denotes the removal of the audio modality and its associated predictor components.

In Table~\ref{ablationresult}, the comparison between (1), (2), and (3) shows that the textual predictor performs the best when the three modality predictors operate independently, as the question is most similar to the subtitle text. Comparing (4) with (7) and (5) with (8) indicates that introducing the audio modality improves performance, as it compensates for missing subtitles in video segments. Additionally, the comparison of (1)~(3) with (6)~(8) demonstrates that the DTL function effectively enhances performance by allowing the modality predictors to learn from one another and mitigate their individual limitations. Finally, comparing (1), (4), and (6) shows that progressively incorporating audio and textual modalities into the visual modality leads to improved performance, reflecting the consistency and complementarity among the modalities.

\subsection{Case Study}
As shown in Figure~\ref{casestudy}, our qualitative analysis demonstrates that incorporating additional modalities enhances the input information and improves performance on the AMVAL. In case (1), where the answer includes actions like "coughing" and "patting," the audio modality leads to better performance compared to visual-based methods. This shows that the audio modality effectively compensates for incomplete information in video segments without subtitles. In case (2), where the video answer involves multiple similar actions, such as "wiping the patient's eyes" and "administering eye drops," the model performs poorly, as the visual and audio modalities introduce noise.

As shown in Figure~\ref{tsne} (a), before modality interaction, features from the three modalities of the same video are scattered, indicating significant differences in the feature space. Notably, the audio features form an almost independent cluster, suggesting the distinctiveness of the audio modality. This structural divergence reflects the complementary nature of the modalities. In Figure~\ref{tsne} (b), after modality interaction, the feature vectors of the three modalities converge, demonstrating the model’s ability to fuse multimodal information and learn consistent representations. However, some dispersed visual and audio features remain, suggesting they retain some independence, further emphasizing their complementarity.

\section{Conclusion}
We proposed the unified Audio-Visual-Textual Span Localization (AVTSL) method to address the challenges in Audio-enhanced Multilingual Video Answer Localization (AMVAL). Our method introduces three predictors based on visual, audio-visual and textual representations, marking the first instance of incorporating the audio modality into the AMVAL. To bridge the gap between modalities, we designed the Audio-Visual-Textual consistency module, incorporating the Dynamic Triangular Loss (DTL) function, which facilitates mutual learning among the modality predictors and fully exploits the consistency and complementarity of multimodal information. Extensive comparative experiments with several state-of-the-art methods demonstrate that our AVTSL achieves superior performance, validating its effectiveness. In future work, we aim to develop larger multilingual medical instructional video question-answering datasets to pre-train multimodal models for AMVAL.

\section{Limitations}
Although AVTSL achieves promising improvements, it still has some limitations. The dataset is limited to medical videos and bilingual (Chinese-English) content, which may hinder generalization to other domains and languages. Additionally, our method targets a single continuous segment. However, in real-world scenarios, answers may consist of multiple disjointed segments, requiring more than one start and end timestamp. This could limit the method's ability to handle complex cases where the relevant information is scattered across different parts of the video.

\bibliography{custom}

\end{document}